\documentstyle[12pt,amsfonts]{article}
\parskip=\medskipamount

\def\appendix#1{
  \addtocounter{section}{1}
  \setcounter{equation}{0}
  \renewcommand{\thesection}{\Alph{section}}
 \section*{Appendix \thesection\protect\indent
 \parbox[t]{11.715cm} {#1}}
 \addcontentsline{toc}{section}{Appendix \thesection\ \ \ #1}
  }
\renewcommand{\thefootnote}{\fnsymbol{footnote}}

\newcommand {\cD}{{\cal D}}

\newcommand {\cF}{{\cal F}}

\newcommand {\cN}{{\cal N}}

\newcommand {\cW}{{\cal W}}

\def\a{\alpha}
\def \bi{\bibitem}

\def\b{\beta}

\def\d{\delta}

\def\G{\Gamma}

\def\l{\lambda}

\def\o{\omega}

\def\q{\theta}
\def\r{\rho}
\def\s{\sigma}
\def\t{\tau}

\def\x{\xi}
\def\z{\zeta}
\def\D{\Delta}
\def\F{\Phi}
\def\J{\Psi}

\def\O{\Omega}

\def\U{\Upsilon}

\newcommand{\ad}{{\dot{\alpha}}}                           
\newcommand{\bd}{{\dot{\beta}}}                            
\newcommand{\ve}{\varepsilon}                            

\newcommand{\pa}{\partial}                           
\newcommand{\hf}{\frac12}

\newcommand{\be}{\begin{equation}}
\newcommand{\ee}{\end{equation}}
\newcommand{\bea}{\begin{eqnarray}}
\newcommand{\eea}{\end{eqnarray}}
\newcommand{\non}{\nonumber}
\begin{document}

\begin{titlepage}
\thispagestyle{empty}

\begin{flushright}
hep-th/0407242 \\
July, 2004
\end{flushright}
\vspace{5mm}

\begin{center}
{\Large\bf  Exact propagators in harmonic superspace
}
\end{center}
\vspace{3mm}

\begin{center}
{\large
Sergei M. Kuzenko 
}\\
\vspace{2mm}

${}$\footnotesize{
{\it
School of Physics, The University of Western Australia\\
Crawley, W.A. 6009, Australia}
} \\
{\tt  kuzenko@physics.uwa.edu.au}
\vspace{2mm}

\end{center}
\vspace{5mm}

\begin{abstract}
\baselineskip=14pt
Within the background field formulation in 
harmonic superspace for quantum 
$\cN=2$ super Yang-Mills theories, 
the propagators of the matter,  gauge and 
ghost superfields possess a complicated 
dependence on the SU(2) harmonic 
variables via the background vector multiplet.
This dependence is shown to simplify drastically 
in the case of an on-shell vector multiplet. For 
a covariantly  constant background 
vector multiplet, we exactly compute 
all the propagators. In conjunction 
with the covariant multi-loop scheme 
developed in hep-th/0302205, these results provide
an efficient (manifestly $\cN=2$ supersymmetric)
technical setup for computing multi-loop 
quantum corrections to effective actions 
in $\cN=2$ supersymmetric gauge theories, 
including the $\cN=4$ super Yang-Mills theory.
\end{abstract}
\vfill
\end{titlepage}

\newpage
\setcounter{page}{1}

\renewcommand{\thefootnote}{\arabic{footnote}}
\setcounter{footnote}{0}

\noindent
Within the background field formulation for 
quantum  $\cN=2$ super Yang-Mills theories 
\cite{BBKO,BKO,BK} in harmonic 
superspace\footnote{The Feynman rules in 
$\cN=2$ harmonic superspace were developed 
in \cite{GIOS1,OY}.}
\cite{GIKOS,GIOS1,GIOS2}, 
there occur three types of background-dependent
Green's  functions:  (i) the $q$-hypermultiplet propagator; 
(ii) the $\o$-hypermultiplet propagator 
(corresponding, in particular, to the Faddeev-Popov 
ghosts); (iii) the vector multiplet propagator. 
Compared with the background-field propagators
in ordinary gauge theories or in $\cN=1$ super Yang-Mills
theories,  the $\cN=2$ superpropagators have a much more 
complicated structure, due to a nontrivial dependence 
on SU(2) harmonics, the internal space variables, 
which enter  even  the corresponding 
covariant  d'Alembertian \cite{BBKO}.
As is  shown below, the harmonic dependence  
of the $\cN=2$ superpropagators simplifies drastically 
if the background vector multiplet satisfies 
classical equations of motion.
Using these observations makes it possible 
to compute the exact propagators, 
in a  straightforward way,  in the presence 
of  a covariantly constant background 
vector multiplet.

We start by assembling the necessary information about  
the $\cN=2$ Yang-Mills supermultiplet
which is known to possess an off-shell formulation 
\cite{GSW} in 
conventional $\cN=2$ superspace ${\Bbb R}^{4|8}$ 
parametrized by coordinates 
$z^A = (x^a, \q^\a_i, {\bar \q}_\ad^i)$, 
where $i =\underline{1},  \underline{2}$. 
The gauge covariant derivatives are defined by  
\be 
\cD_A = (\cD_a, \cD^i_\a, {\bar \cD}^\ad_i )
= D_A + {\rm i}\,  \G_A (z)~,
\ee 
with $D_A = (D_a, D^i_\a, {\bar D}^\ad_i )$
the flat covariant derivatives, and $\G_A$
the gauge connection.
Their gauge transformation law is 
\be
\cD_A ~\to ~ {\rm e}^{{\rm i} \t(z)} \, \cD_A\,
{\rm e}^{-{\rm i} \t(z)}~, \qquad 
\t^\dagger = \t ~, 
\label{tau}
\ee
with the gauge parameter $\t(z)$ being arbitrary modulo 
the reality condition imposed.
The gauge covariant derivatives
obey the algebra \cite{GSW}:
\bea 
\{\cD^i_\a , {\bar \cD}_{\ad j} \}
&=& - 2{\rm i}\,\d^i_j \cD_{\a \ad} \non \\
\{\cD^i_\a , \cD^j_\b \} = 2{\rm i}\, \ve_{\a\b}
\ve^{ij} {\bar \cW}~, \quad && \quad 
\{ {\bar \cD}_{\ad i} , {\bar \cD}_{\bd j} \} 
= 2{\rm i}\, \ve_{\ad \bd} \ve_{ij} \cW~, \non \\
\left[ \cD^i_\a , \cD_{\b \bd} \right]  
= \ve_{\a\b} {\bar \cD}^i_\bd {\bar \cW}~,
 \quad && \quad
[ {\bar \cD}_{\ad i}, \cD_{\b \bd}] 
=\ve_{\ad \bd}\cD_{\b i} \cW ~, \non \\
\left[ \cD_{\a \ad} , \cD_{\b \bd} \right]
= {\rm i}\, \cF_{\a \ad,\b \bd} &=& 
{ {\rm i}\, \over 4} \,  \ve_{\ad \bd}\, 
\cD^i{}_{(\a} \cD_{\b )i} \cW 
- { {\rm i} \over 4}\, \ve_{\a \b}\, 
{\bar \cD}^i{}_{(\ad} {\bar \cD}_{\bd )i} {\bar \cW} ~.
\label{algebra}
\eea
The superfield strengths $\cW$ and $\bar \cW$ satisfy 
the Bianchi identities 
\be
{\bar \cD}^i_\ad \cW = \cD^i_\a {\bar \cW} =0~,
\qquad 
\cD^{ij} \cW = {\bar \cD}^{ij} {\bar \cW}~,
\ee
where 
\be
\cD^{ij} = \cD^{\a (i} \cD_\a^{j)} ~, \qquad 
 {\bar \cD}^{ij} 
= {\bar \cD}_\ad^{(i} {\bar \cD}^{j) \ad}~.
\ee

The $\cN=2$ harmonic superspace ${\Bbb R}^{4|8}\times S^2$
\cite{GIKOS,GIOS1,GIOS2}
extends conventional superspace
by the two-sphere $S^2 ={\rm SU(2)/U(1)}$
parametrized by harmonics, i.e., group
elements
\bea
({u_i}^-\,,\,{u_i}^+) \in {\rm SU}(2)~, \quad
u^+_i = \ve_{ij}u^{+j}~, \quad \overline{u^{+i}} = u^-_i~,
\quad u^{+i}u_i^- = 1 ~.
\eea
In harmonic superspace,
both the $\cN=2$ Yang-Mills supermultiplet and
hypermultiplets can be described 
by {\it unconstrained} 
superfields over the analytic
subspace of ${\Bbb R}^{4|8}\times S^2$
parametrized by the variables
$ \x \equiv (y^a,\q^{+\a},{\bar\q}^+_{\dot\a}, \,
u^+_i,u^-_j) $, 
where the so-called analytic basis is defined by
\be
y^a = x^a - 2{\rm i}\, \q^{(i}\s^a {\bar \q}^{j)}u^+_i u^-_j~, 
\qquad
 \q^\pm_\a=u^\pm_i \q^i_\a~, \qquad {\bar \q}^\pm_{\dot\a}
=u^\pm_i{\bar \q}^i_{\dot\a}~.
\ee
With the notation 
\be 
 \cD^\pm_\a=u^\pm_i \cD^i_\a~, 
\qquad {\bar \cD}^\pm_{\dot\a}
=u^\pm_i{\bar \cD}^i_{\dot\a}~,
\ee
it follows from (\ref{algebra}) that 
the operators $\cD^+_\a$ and 
${\bar \cD}^+_{\dot\a}$ strictly anticommute, 
$ \{ \cD^+_\a , \cD^+_\b \} =
\{ {\bar \cD}^+_{\dot\a} , {\bar \cD}^+_{\dot\b} \}
= \{ \cD^+_\a , {\bar \cD}^+_{\dot\a} \} =0$.
A covariantly analytic superfield
$\F^{(p)}(z,u)$ is defined to 
obey  the constraints
\be
\cD^+_\a \F^{(p)} = {\bar \cD}^+_\ad \F^{(p)}=0~. 
\label{analytic-con}
\ee
Here the superscript $p$ refers to the harmonic 
U(1) charge, $\cD^0 \, \F^{(p)} = p\,  \F^{(p)}$, 
where $\cD^0$ is one of the harmonic gauge 
covariant  derivatives\footnote{Throughout this 
paper, we use the so-called $\t$-frame 
\cite{GIKOS,GIOS2} defined in  the appendix.}
\be 
\cD^0 = u^{+i} \, \frac{\pa} {\pa u^{+i}} 
- u^{-i} \, \frac{\pa} {\pa u^{-i}} ~,
\qquad
\cD^{\pm \pm} = u^{\pm i} \, 
\frac{\pa} {\pa u^{\mp i}}  ~.
\ee
The operator $\cD^{++}$ acts on the space 
of covariantly analytic superfields.

In the framework of the background field 
formalism in $\cN=2 $ harmonic superspace
\cite{BBKO,BKO,BK}, 
there appear three types of covariant 
(matter and gauge field) propagators:
\bea
G^{(1,1)} (z,u
\, \mbox{{\bf ,}} \, z',u')
&=& \phantom{-}
\frac{1}{{\stackrel{\frown}{\Box}}{}}\,
(\cD^+)^4 \,(\cD'^+)^4\,
\Big\{ {\bf 1}\, \delta^{12}(z-z')\,
{1 \over (u^+ u'^+)^3}\Big\}~, \non \\
G^{(0,0)} (z,u
\, \mbox{{\bf ,}} \,  z',u')
&=& -\frac{1}{{\stackrel{\frown}{\Box}}{}}\,
(\cD^+)^4 \,(\cD'^+)^4\,
\Big\{ {\bf 1}\, \delta^{12}(z-z')\,
{(u^-u'^-) \over (u^+ u'^+)^3}\Big\}~,\\
G^{(2,2)} (z,u
\, \mbox{{\bf ,}} \, z',u')
&=& -\frac{1}{{\stackrel{\frown}{\Box}}}\,
(\cD^+)^4 
\Big\{ {\bf 1}\, \delta^{12}(z-z')\,
\d^{(-2,2)} (u,u')\Big\}~. \non
\eea
Here  ${\stackrel{\frown}{\Box}}$ is 
the analytic d'Alembertian \cite{BBKO},
\bea
{\stackrel{\frown}{\Box}}{}&=&
{\cal D}^a{\cal D}_a -
\frac{{\rm i}}{2}({\cal D}^{+\a}\cW){\cal D}^-_\a
-\frac{{\rm i}}{2}
({\bar{\cal D}}^+_{\dot\alpha}{\bar \cW}){\bar{\cal D}}^{-{\dot\alpha}}
+\frac{{\rm i}}{4}({\cal D}^{+\a} {\cal D}^+_\a \cW) \cD^{--}\non \\
&{}& -\frac{{\rm i}}{8}[{\cal D}^{+\alpha},{\cal D}^-_\alpha] \cW
- \frac{1}{2}\{{\bar \cW},\cW \}~,
\label{analdal}
\eea
$ \delta^{12}(z-z')$ denotes the $\cN=2$ superspace 
delta-function, 
\be
\delta^{12}(z-z') = \d^4(x-x')\, (\q-\q')^4 
({\bar \q} -{\bar \q}')^4~,
\ee
 $(u^+ u'^+)^{-3}$ and $(u^-u'^-)(u^+ u'^+)^{-3}$
denote special  harmonic distributions \cite{GIOS1,GIOS2},
and finally  the two-point function
$\d^{(p,-p)} (u,u')$, with $p \in {\Bbb Z}$,  
stands for a harmonic 
delta-function  \cite{GIOS1,GIOS2}.
The Green's function $G^{(1,1)}$ 
determines the Feynman propagator
of a covariantly analytic matter superfield
$q^+$ (the so-called $q$-hypermultiplet \cite{GIKOS,GIOS2}) 
transforming in some representation $R$ of the gauge group,
\be  
{\rm i} \,\langle 
{ q}^+(z,u)\,\breve{ q}{}^+(z',u') \rangle 
= G^{(1,1)} (z,u \, \mbox{{\bf ,}} \, z',u')~,
\ee
with $\breve{ q}{}^+ $ the analyticity-preserving 
conjugate of $q^+$. 
The Green's function $G^{(0,0)}$ 
determines the Feynman propagator
of a covariantly analytic matter (or, in the case of 
the adjoint representation,   ghost) superfield
$\o$ (the so-called $\o$-hypermultiplet \cite{GIKOS,GIOS2}) 
in the representations $R$ of the gauge group,
\be  
{\rm i} \,\langle 
\o(z,u)\,\breve{ \o}{}(z',u') \rangle 
= G^{(0,0)} (z,u \, \mbox{{\bf ,}} \, z',u')~,
\ee
When $R$ coincides with the adjoint representation, 
the Green's function $G^{(2,2)}$  determines the 
Feynman propagator of the quantum gauge superfield, 
\be  
{\rm i} \,\langle 
v^{++}(z,u)\, v^{++}{}^{\rm T}(z',u') \rangle 
= G^{(2,2)} (z,u \, \mbox{{\bf ,}} \, z',u')~,
\ee
see \cite{BBKO,BKO,BK} for more details. 
In what follows, we will study in detail only 
the Green's functions $G^{(1,1)}$ and 
$G^{(2,2)}$, since the structure of $G^{(0,0)}$  
is very similar to that of $G^{(1,1)}$. 

The  operator ${\stackrel{\frown}{\Box}}$
acts on the space of covariantly analytic superfields.
${}$Given such a superfield $\F^{(p)}$
obeying the constraints
(\ref{analytic-con}),  one can check 
\be
\hf (\cD^+)^4 (\cD^{--})^2 \, \F^{(p)} 
=  {\stackrel{\frown}{\Box}}\, \F^{(p)}~.
\ee
Among the important properties of
${\stackrel{\frown}{\Box}}$ is 
the following \cite{BKO}:
\bea 
(\cD^+)^4 \, {\stackrel{\frown}{\Box}}
&=& {\stackrel{\frown}{\Box}} \, (\cD^+)^4~. 
\label{prop1}
\eea
In addition, for an arbitrary  superfield 
$V^{(p)}(z,u)$ of U(1) charge $p$, 
we have
\be
[ \cD^{++},  {\stackrel{\frown}{\Box} }]\, V^{(p)}
= { {\rm i} \over 2}
\Big\{ \hf
(p-1) \,(\cD^+\cD^+ \cW)
- ({\cal D}^{+\a}\cW){\cal D}^+_\a
- ({\bar{\cal D}}^+_{\dot\alpha}{\bar \cW})
{\bar{\cal D}}^{+{\dot\alpha}}
\Big\} V^{(p)}~.
\label{prop2}
\ee
Eq. (\ref{prop1}) implies, in particular, 
that $G^{(1,1)}$ and $G^{(2,2)}$
can be rewritten as follows:
\bea
G^{(1,1)} (z,u
\, \mbox{{\bf ,}} \, z',u')
&=& 
\phantom{-}
(\cD^+)^4 \,(\cD'^+)^4 \,
\frac{1}{{\stackrel{\frown}{\Box}}{}}\,
\Big\{
{\bf 1}\, \delta^{12}(z-z')\,
{1 \over (u^+ u'^+)^3}\Big\}~, \non \\
G^{(2,2)} (z,u
\, \mbox{{\bf ,}} \, z',u')
&=& - (\cD^+)^4 \, 
\frac{1}{{\stackrel{\frown}{\Box}}}\,
\Big\{ {\bf 1}\, \delta^{12}(z-z')\,
\d^{(-2,2)} (u,u')\Big\}~. 
\label{propagators}
\eea

Sometimes, it is useful to rewrite $G^{(2,2)} $
in  a manifestly analytic form, 
following  \cite{GIOS2}, 
\bea
G^{(2,2)} (z,u
\, \mbox{{\bf ,}} \, z',u')
&=& - \frac{1}{2 \,{\stackrel{\frown}{\Box}}{}^2 }\,
(\cD^+)^4  (\cD'^+)^4  
\Big\{ {\bf 1}\, \delta^{12}(z-z')\,
(\cD^{--})^2 \d^{(2,-2)} (u,u')\Big\}~. 
\eea
This representation may be, in principle,
advantageous
when handling those supergraphs which 
contain a product of harmonic distributions
and require the introduction of a harmonic 
regularization at intermediate stages of 
the calculation.

Let us introduce a new second-order operator $\D$, 
\be 
\D = {\stackrel{\frown}{\Box} }
 + \frac{{\rm i}}{2}({\cal D}^{-\a}\cW){\cal D}^+_\a
+\frac{{\rm i}}{2}
({\bar{\cal D}}^-_\ad{\bar \cW}){\bar{\cal D}}^{+\ad}~,
\ee
which coincides with ${\stackrel{\frown}{\Box} }$
on the space of covariantly analytic superfields, 
\be
\cD^+_\a \F^{(p)} = {\bar \cD}^+_\ad \F^{(p)}=0 \quad 
\Longrightarrow \quad 
 \D\, \F^{(p)} =
{\stackrel{\frown}{\Box} } \, 
\F^{(p)} ~.
\ee
In terms of the operator introduced, 
the property (\ref{prop1}) turns into 
\be
(\cD^+)^4 
\Big\{ 
\D +{ {\rm i} \over 2}\,
[{\cal D}^{+\alpha},{\cal D}^-_\alpha] \cW 
\Big\} 
= \D \, (\cD^+)^4~,
\ee
while eq. (\ref{prop2}) turns into 
\be
[ \cD^{++},  \D ]\, V^{(p)}
= { {\rm i} \over 4} (p-1) \,(\cD^+\cD^+ \cW)
V^{(p)}~.
\ee

If the background vector multiplet satisfies 
the classical equations of motion, 
\be 
\cD^{ij} \cW = {\bar \cD}^{ij} {\bar \cW} = 0~,
\label{on-shell}
\ee
then the operator $\D$ is characterized 
by the following two properties:
\be 
(\cD^+)^4 \, \D 
= \D \, (\cD^+)^4~,\qquad \quad 
[ \cD^{++},  \D ] = 0~.
\ee
The latter is equivalent to the fact that 
$\D$ is harmonic-independent,\footnote{The
operator $\D$ could have been introduced 
several years ago in \cite{BK}.} 
\be 
\D = \cD^a \cD_a 
+ \frac{{\rm i}}{2}
(\cD^\a_i \cW)\cD_\a^i
-\frac{{\rm i}}{2}
({\bar \cD}^i_\ad{\bar \cW}){\bar \cD}^\ad_i
- \frac{1}{2}\{{\bar \cW},\cW \}~,
\label{Delta}
\ee
while it maps the space of covariantly anaytic
superfields into itself.
Taking into account the properties of $\D$ described, 
the Green's functions 
$G^{(1,1)} $ and $G^{(2,2)} $
can be rewritten in the form:
\bea
G^{(1,1)} (z,u
\, \mbox{{\bf ,}} \, z',u')
&=& 
\phantom{-}
(\cD^+)^4 \,(\cD'^+)^4\,
\frac{1}{\D}\,
\Big\{
{\bf 1}\, \delta^{12}(z-z') \Big\}\,
{1 \over (u^+ u'^+)^3}~, \non \\
G^{(2,2)} (z,u
\, \mbox{{\bf ,}} \, z',u')
&=& - (\cD^+)^4 \, 
\frac{1}{\D}\,
\Big\{ {\bf 1}\, \delta^{12}(z-z') \Big\} \,
\d^{(-2,2)} (u,u') 
\label{propagators2} \\
&=& - (\cD^+)^4 (\cD'^+)^4   \, 
\frac{1}{2\D^2}\,
\Big\{ {\bf 1}\, \delta^{12}(z-z') \Big\} \,
(\cD^{--})^2 \,\d^{(2,-2)} (u,u')~. \non
\eea
The dependence of $G^{(1,1)} $ and
$G^{(2,2)} $ on the harmonic is completely 
factorized.\footnote{This technical result 
is actually of utmost importance. 
The point is that it allows us to keep 
the harmonic dependence of 
$\cN=2$ supergraphs under control.
Some supergraphs  may
involve potentially dangerous coinciding 
 harmonic singularities \cite{GIOS2,BK,KM4} 
which have to be treated extremely carefully.}
At this stage, it is advantageous to introduce, 
following Fock and Schwinger, 
the proper-time representation:
\bea
 \frac{1}{\D^n}\,
\Big\{
{\bf 1}\, \delta^{12}(z-z') \Big\}
&=& \frac{ (-{\rm i})^n}{(n-1)!}  
\int_{0}^{\infty} {\rm d}s \, s^{n-1}\,
K(z,z'|s) \, {\rm e}^{-\ve \, s}~, 
\qquad 
\ve \to +0 \non \\
K(z,z'|s) &=& {\rm e}^{{\rm i} s \D} \,
\Big\{ {\bf 1} \, \d^{12}(z-z')\Big\} ~,
\label{proper-time}
\eea
with $n=1$ or 2.

A further simplification occurs in the case of 
a covariantly constant vector multiplet, 
\be 
\cD_a \cW = \cD_a {\bar \cW} =0 
\quad \Longrightarrow \quad 
[\cW, {\bar \cW}] =0~.
\ee
Then, the first-order operator appearing in  (\ref{Delta}), 
\be 
\U = \frac{{\rm i}}{2}
(\cD^\a_i \cW)\cD_\a^i
-\frac{{\rm i}}{2}
({\bar \cD}^i_\ad{\bar \cW}){\bar \cD}^\ad_i             
-{\bar \cW} \cW ~,
\ee
turns out to commute with the vector covariant derivative,
\be 
[\U , \cD_a] = 0~,
\ee
and this is similar to the $\cN=1$ case
\cite{KM1}. As in the $\cN=1$ case, 
we can now associate with the heat  
kernel $K(z,z'|s)$,  eq. (\ref{proper-time}), 
a reduced kernel $\tilde{K}(z,z'|s)$. 
The latter is defined as  follows:
\bea 
K(z,z'|s) &=& 
= {\rm e}^{{\rm i} s \U} \,
{\rm e}^{{\rm i} s \cD^a \cD_a} \,
\Big\{ {\bf 1} \, \d^{12}(z-z') \Big\}
={\rm e}^{{\rm i} s \U} \,
\tilde{K}(z,z'|s)~.
\eea

The reduced heat kernel $\tilde{K}(z,z'|s)$
can now be evaluated 
in the same way as it was done in 
the $\cN=1$ superspace case \cite{Ohr1,KM1}
by generalizing the Schwinger construction
\cite{Schwinger}.
The resullt is
\be
\tilde{K}(z,z'|s) = -\frac{\rm i}{(4 \pi s)^2} \, 
\det{}^{1/2} \left( \frac{ s \cF}
{\sinh  ( s \cF) }
\right)
\, {\rm e}^{ \frac{{\rm i}}{4} \r^a (\cF \coth (s \cF))_{ab} \r^b}
\, \z^4 \, \bar{\z}^4 \,I(z,z')
~,
\label{Ksol}
\ee
where the determinant is computed with respect 
to the Lorentz indices, 
and 
\be 
\z^4 = (\q-\q')^4~, \qquad 
{\bar \z}^4 = ({\bar \q}-{\bar \q}')^4~.
\ee
Here we have introduced the $\cN=2$ 
supersymmetric  interval
$\z^A \equiv \z^A (z,z') = -\z^A (z',z) $ 
defined by
\bea
 \z^A = 
\left\{
\begin{array}{l}
\r^a = (x-x')^a - {\rm i} (\q-\q')_i \s^a {\bar \q}'^i 
+ {\rm i} \q'_i \s^a ( {\bar \q} - {\bar \q}')^i ~, \\
\z^\a_i = (\q - \q')^\a_i ~, \\
{\bar \z}_\ad^i =({\bar \q} -{\bar \q}' )_\ad^i ~. 
\end{array} 
\right. 
\label{super-interval}
\eea
The parallel displacement propagator, 
$I(z,z') $, and its properties \cite{KM1} are collected 
in the appendix.
The reduced kernel turns out to
solve the evolution problem 
\be 
\Big( {\rm i}\, { {\rm d}\over {\rm d}s} 
+ \cD^a \cD_a \Big)\, \tilde{K}(z,z'|s) = 0~, 
\qquad 
\tilde{K}(z,z'|0 ) = \d^{12}(z-z')
=\d^4 (\r)\, \z^4\,{\bar \z}^4
\ee
as a consequence, in particular,  
of eq. (\ref{super-PDO3}) and 
of the identity 
\be
  \z^4 \, \bar{\z}^4 \, \cD_a I(z,z') 
= -\, \frac{{\rm i}}{2} \,  \z^4 \,
\bar{\z}^4 \,  \cF_{ab} \,\r^b  I(z,z')~,
\ee
which  the parallel displacement propagator
obeys, see the appendix. 

In order to obtain the complete kernel $K(z,z'|s)$, 
one still has to evaluate the action of the operator
$\exp ({\rm i} s \U)$ on the reduced kernel 
$\tilde{K}(z,z'|s) $.  This is accomplished 
in complete analogy with the $\cN=1$ case
\cite{KM1}
\bea
K(z,z'|s) &=& -\frac{\rm i}{(4 \pi s)^2} \, 
\det{}^{1/2} \left( \frac{ s \cF}
{\sinh  ( s \cF) }
\right)
\, {\rm e}^{ \frac{{\rm i}}{4} \r(s)
 \cF \coth (s \cF) \r (s) }
\, \z^4 (s) \, \bar{\z}^4(s) \,
 {\rm e}^{{\rm i} s \U} I(z,z')~, \non \\
\z^A(s) &= &{\rm e}^{{\rm i} s \U}  \, \z^A \,
{\rm e}^{-{\rm i} s \U} ~.
\eea 
The components of 
$\z^A(s) = (\r^a (s), \z^\a_i (s), {\bar \z}^i_\ad (s) ) $
can be easily evaluated using the (anti) commutation
relations (\ref{algebra}) and   
the obvious identity
\be 
\cD_B \, \z^A = \d_B{}^A 
+\hf \, \z^C \, T_{CB}{}^A~,
\label{trivial}
\ee
with $T_{CB}{}^A$ the flat superspace torsion.
The action of $\exp ({\rm i} s \U)$ on the 
parallel displacement propagator $I(z,z') $
is evaluated using the relation (\ref{super-PTO-der1})
or (\ref{super-PTO-der-mod}); the result is quite 
lengthy and is not reproduced here.

Making use of the (anti)commutation relations for the
covariant derivatives, eq. (\ref{algebra}), 
 one can check that 
\bea
\left[\D, \cD^i_\a \right] 
= - {{\rm i}\over 2}\, \cD^j{}_{(\a}\cD_{\b) j} \cW \, 
\cD^{\b i}~, \qquad 
\left[\D, {\bar \cD}^i_\ad \right] 
={{\rm i}\over 2}\,
{\bar \cD}^j{}_{(\ad}{\bar \cD}_{\bd) j} {\bar \cW} \,
{\bar \cD}^{\bd i}~, 
\eea
and therefore 
\be 
\left[\D, \cD^{ij} \right] =0~, 
\qquad
\left[\D, {\bar \cD}^{ij} \right] =0~.
\ee
The latter identities imply 
the following important properties:
\bea
\cD_{ij} \, K(z,z'|s) = \cD'_{ij} \, K(z,z'|s)~,
\qquad 
{\bar \cD}_{ij} \, K(z,z'|s) 
= {\bar \cD}'_{ij} \, K(z,z'|s)~.
\label{shift}
\eea

Since 
\be
(\cD^+)^4 = {1 \over 16} \, \cD^{ij} \, 
{\bar \cD}^{kl} \, u^+_i u^+_j u^+_k u^+_l
= {1 \over 16} \, {\bar \cD}^{ij} \, 
\cD^{kl} \, u^+_i u^+_j u^+_k u^+_l
 ~,
\ee
the relations (\ref{shift}) allow one
to accumulate the overall D-factors, 
which occur in 
\be
(\cD^+)^4 (\cD'^+)^4\, K(z,z'|s)
\ee
at different superspace points, 
$z$ and $z'$, to a single point, say, $z$.
One then obtains
\bea
(\cD^+)^4 (\cD'^+)^4\, K(z,z'|s) &=& (\cD^+)^4
\Big\{ (u^+u'^+)^4 \,(\cD^-)^4\, 
-{{\rm i}\over 2} (u^+u'^+)^3(u^-u'^+)\, \O^{--} \non \\
&& \qquad \qquad  + (u^+u'^+)^2(u^-u'^+)^2\,\D \Big\} K(z,z'|s)~,
\label{crucial}
\eea
where 
\be
\O^{--} = \cD^{\a\ad} \cD^-_\a \cD^-_\ad 
+\hf \cW (\cD^-)^2 +\hf {\bar \cW}({\bar \cD}^-)^2
+(\cD^-\cW)\cD^- + ({\bar \cD}^-{\bar \cW}){\bar \cD}^-~.
\ee
The relation (\ref{crucial}) is analogous to 
the representation for the two-point function 
\be 
(\cD^+)^4 (\cD'^+)^4 \Big\{ {\bf 1}\,  \d^{12}(z-z') \Big\}
\ee
derived in \cite{KM2}. 
The last term in eq. (\ref{crucial})
can be rewritten in an equivalent form by 
taking into account the fact that $K(z,z'|s)$ 
satisfies the evolution equation 
\be 
\Big( {\rm i}\, { {\rm d}\over {\rm d}s} 
~+~ \D \Big)\, K(z,z'|s)~ =~ 0~.
\ee

It also follows from eq. (\ref{shift}) 
and the obvious  property of harmonic delta functions
\be
 u^+_i u^+_j u^+_k u^+_l  \,  \d^{(-2,2)}(u,u')
=  u'^+_i u'^+_j u'^+_k u'^+_l  \,  \d^{(2,-2)}(u,u')
\ee
that 
\be
(\cD^+)^4 \Big\{ K(z,z'|s) \, \d^{(-2,2)}(u,u')\Big\} 
=(\cD'^+)^4  
 \Big\{ K(z,z'|s) \, \d^{(2,-2)}(u,u')\Big\} ~.
\ee
The latter guarantees that the gauge field  Green's function
$G^{(2,2)}(z,u;z',u')$ is 
covariantly analytic in both arguments. 

The action of a product of covariant derivatives
on the kernel $K(z,z'|s) $ can be readily evaluated 
using  the algebra of covariant derivatives (\ref{algebra})
and 
the fundamental relations (\ref{super-PTO-der1})
and (\ref{super-PTO-der-mod}).

In conclusion, we would like to list several interesting 
open problems that can be addressed using 
the approach advocated in this note.  
First of all, it becomes  now feasible  to compute, 
in a manifestly $\cN=2$ supersymmetric way,  
multi-loop quantum corrections to the effective 
action for a non-Abelian vector multiplet which 
is only constrained to be on-shell but otherwise arbitrary.
Indeed, the representation (\ref{propagators2})
implies that the harmonic dependence of the 
propagators is completely factorized (this dependence 
is actually of the same form possessed
by  the free propagators), 
and therefore under control. 
As regards the superfield heat kernel 
$K(z,z'|s)$,  eq. (\ref{proper-time}), 
its manifestly  $\cN=2$  supersymmetric  
and gauge covariant derivative expansion 
can be developed in complete analogy with 
the $\cN=1$ case worked out in \cite{KM1}.
Another application is the calculation 
of low-energy effective actions for  $\cN=2,4$ 
super Yang-Mills theories on the Coulomb branch, 
including $\cN=2$ supersymmetric 
Heisenberg-Euler\footnote{See \cite{Dunne}
for a recent review of Heisenberg-Euler 
effective Lagrangians.} 
type actions \cite{BKT}. Calculation of
multi-loop  quantum corrections to such actions 
can be carried out with  the use
of the exact propagators, in the presence of 
a covariantly constant vector multiplet, 
which have been found in this note.

It has recently been  demonstrated 
\cite{DSch}  (see also \cite{Dunne} 
and references therein)
that for a self-dual background
the two-loop QED effective action 
takes a remarkably simple form 
that is very similar to the one-loop action 
in the same background.
There are expectations that 
this  similarity persists at higher loops, 
and therefore there should be some remarkable 
structure encoded in the all-loop effective action 
for gauge theories. In the supersymmetric case, 
one has to replace the requirement of self-duality 
by that of relaxed super self-duality  
\cite{KM5} in order to arrive at conclusions
similar to those given in \cite{DSch}.  Further 
progress in this direction may be 
achieved through the analysis of
$\cN=2$ covariant supergraphs. 

${}$Finally, we believe that the results of this note may 
be helpful in the context of 
the conjectured correspondence 
\cite{Maldacena,CT,BPT}
between the D3-brane action in $AdS_5 \times S^5$
and the low-energy action for $\cN=4$ SU$(N)$ 
SYM on its Coulomb branch,  with the gauge group 
SU$(N)$ spontaneously broken to 
SU$(N-1) \times {\rm U}(1)$. 
There have appeared two independent 
$F^6$ tests of this conjecture \cite{BPT,KM6}, 
with conflicting conclusions.
The approach advocated here provides the 
opportunity for a further test. 
\vskip.5cm

\noindent
{\bf Acknowledgements:}\\
E-mail correspondence with Nikolay Pletnev, 
and discussions with Ian McArthur and 
Christian Schubert are gratefully acknowledged.
Special thanks are due to Ian McArthur 
for reading the manuscript.
This work was  completed during the author's visit 
in June--July, 2004
to the Max Planck Institute for Gravitational Physics
(Albert Einstein Institute), Golm. 
Hospitality of Stefan  Theisen and the 
Max Planck Society is gratefully acknowledged.
This work is supported in part by the Australian Research
Council. 
\renewcommand{\theequation}{\thesection.\arabic{equation}}

\begin{appendix}
{Parallel displacement propagator}
In this appendix we describe, basically following \cite{KM1}, 
the main properties of the 
parallel displacement propagator
$I(z,z')$ in $\cN=2$ superspace. 
This object is uniquely specified by 
the following requirements:\\
(i) the gauge transformation law
\be
 I (z, z') ~\to ~
{\rm e}^{{\rm i} \t(z)} \,  I (z, z') \,
{\rm e}^{-{\rm i} \t(z')} ~
\label{super-PDO1}
\ee
with respect to the  gauge ($\t$-frame) transformation 
of  the covariant derivatives (\ref{tau});\\
(ii) the equation  
\be
\z^A \cD_A \, I(z,z') 
= \z^A \Big( D_A +{\rm i} \, \G_A(z) \Big) I(z,z') =0~;
\label{super-PDO2}
\ee
(iii) the boundary condition 
\be 
I(z,z) ={\bf 1}~.
\label{super-PDO3}
\ee
These imply the important relation
\be
I(z,z') \, I(z', z) = {\bf 1}~,
\label{collapse}
\ee
as well as 
\be
\z^A \cD'_A \, I(z,z') 
= \z^A  \Big( D'_A \,I(z,z') 
 - {\rm i} \, I(z,z') \, \G_A(z') \Big) =0~.
\ee

Let $\J(z)$ be a {\it harmonic-independent}
superfield transforming in some representation of 
the gauge group, 
\be 
\J(z) ~\to ~ {\rm e}^{{\rm i} \t(z)} \, \J(z)~.
\ee
Then it can be represented by 
the  covariant Taylor series \cite{KM1} 
\be 
\J(z) = I(z,z') \,\sum_{n=0}^{\infty} {1 \over n!}\,
\z^{A_n} \ldots \z^{A_1} \, 
\cD'_{A_1} \ldots \cD'_{A_n} \, \J(z') ~. 
\label{super-Taylor2}
\ee

The fundametal properties of the
parallel displacement propagator are 
\cite{KM1}
\bea 
\cD_B I(z,z') &=& {\rm i} \, I(z,z') \,
\sum_{n=1}^{\infty} { 1  \over (n+1)!} \,
\Big\{
n \, \z^{A_n} 
\ldots \z^{A_1}  
\cD'_{A_1} \ldots \cD'_{A_{n-1} } \cF_{A_n \,B } (z') 
\label{super-PTO-der1} \\
&+& \hf  
(n-1)\, 
\z^{A_n} T_{A_n \,B}{}^C \,\z^{A_{n-1}} 
\ldots \z^{A_1}  
\cD'_{A_1} \ldots \cD'_{A_{n-2} } \cF_{A_{n-1} \,C } (z') \Big\}~,
\non 
\eea 
and
\bea 
\cD_B I(z,z') &=& {\rm i} \,
\sum_{n=1}^{\infty} {(-1)^{n} \over (n+1)! } \, \Big\{
- \z^{A_n} 
\ldots \z^{A_1}  
\cD_{A_1} \ldots \cD_{A_{n-1} } \cF_{A_n \,B } (z) 
\label{super-PTO-der-mod} \\
&+& \hf  
(n-1)\,
\z^{A_n} T_{A_n \,B}{}^C \,\z^{A_{n-1}} 
\ldots \z^{A_1}  
\cD_{A_1} \ldots \cD_{A_{n-2} } \cF_{A_{n-1} \,C } (z) \Big\}
\, I(z,z')~.
\non 
\eea
Here $\cF_{AB}$ denotes the superfield strength defined
as follows
\be
[\cD_A , \cD_B \} = T_{AB}{}^C \cD_C 
+{\rm i}\, \cF_{AB}~.
\ee
In the case of a covariantly constant vector multiplet, 
the series in (\ref{super-PTO-der1}) and
(\ref{super-PTO-der-mod}) terminate as only the tensors
$\cF_{CD}$, $\cD_A\cF_{CD}$ and $\cD_A \cD_B \cF_{CD}$
have  non-vanishing components.

Throughout this paper, we work in the
$\t$-frame, in which the gauge covariant derivatives $\cD_A$ 
are harmonic-independent, 
although the harmonic superspace practitioners  often use
the so-called $\l$-frame \cite{GIKOS,GIOS2}. 
To go over to the $\l$-frame, one has to transform 
\be
\cD_A \quad \longrightarrow \quad {\rm e}^{{\rm i}\O} \, \cD_A \,
{\rm e}^{-{\rm i}\O}~,
\ee 
and similarly for matter superfields, 
where $\O(z,u) $ is the bridge superfield 
\cite{GIKOS,GIOS2} which occurs as follows
\be 
\cD^+_\a = {\rm e}^{-{\rm i}\O} \, D^+_\a \,
{\rm e}^{{\rm i}\O}~, 
\qquad 
{\bar \cD}^+_\a = {\rm e}^{-{\rm i}\O} \, 
{\bar D}^+_\a \,
{\rm e}^{{\rm i}\O}~.
\ee
In the $\l$-frame, the gauge covariant derivative 
$\cD^+_a$ and ${\bar \cD}^+_\ad$ coincide 
with the flat derivatives $D^+_\a$ and 
${\bar D}^+_\ad$, respectively. 
The parallel displacement propagator
in the $\l$-frame is obtained from that in 
the $\t$-frame by the transformation 
\be
I(z,z') \equiv I_\t(z,z') 
\quad \longrightarrow \quad 
{\rm e}^{{\rm i}\O(z,u)} \, I(z,z') \,
{\rm e}^{-{\rm i}\O(z',u')}
\equiv I_\l(z,u  \, \mbox{{\bf ,}} \, z',u')
~.
\ee 

\end{appendix}

\end{document}